\documentclass{svjour3}                     
\smartqed  
\usepackage{epsfig}
\usepackage{natbib}

\usepackage{amssymb}

\newcommand{\be}{\begin{equation}}
\newcommand{\ee}{\end{equation}}
\newcommand{\ba}{\begin{array}}
\newcommand{\ea}{\end{array}}

\begin{document}

\title{Modelling radiation-induced cell cycle delays
\thanks{This work was supported in part (A. O-M) by the grant of Polish
State Committee for Scientific Research (KBN, Grant No 1 P03B 159
29). Moreover, E. G-N acknowledges Marie Curie TOK COCOS grant at
the Mark Kac Complex Systems Research Center in Krak\'ow, Poland.
E.N. was supported by BMBF (Bonn, Germany) under contract number
02S8203 and 02S8497.}
}


\author{Anna Ochab-Marcinek         \and
        Ewa Gudowska-Nowak   \and
    Elena Nasonova      \and
    Sylvia Ritter
}


\institute{A. Ochab-Marcinek \at
              Department of Soft Condensed Matter, Institute of Physical Chemistry, 
Polish Academy of Sciences, ul. Kasprzaka 44/52, Warsaw, Poland \\
              \email{ochab@ifka.ichf.edu.pl}           
            \and
           E. Gudowska-Nowak \at
             Marian Smoluchowski Institute of Physics, Jagiellonian University, ul. Reymonta 4, Krak{\'o}w, Poland\\
        Mark Kac Complex Systems Research Centre, Jagiellonian University, Reymonta 4, Krak{\'o}w, Poland\\
        Biophysics Department, Gesellschaft f{\"u}r Schwerionenforschung (GSI),
Planckstrasse 1, Darmstadt, Germany
%
\and
    E. Nasonova \at
    Biophysics Department, Gesellschaft f{\"u}r Schwerionenforschung (GSI),
 Planckstrasse 1, Darmstadt, Germany\\
 Joint Institute for Nuclear Research (JINR),
 141980 Dubna, Moscow Region, Russia
 \and
    S. Ritter \at
    Biophysics Department, Gesellschaft f{\"u}r Schwerionenforschung (GSI),
 Planckstrasse 1, Darmstadt, Germany
}

\date{Received: date / Accepted: date}

\maketitle

\begin{abstract}
Ionizing radiation is known to delay the cell cycle progression. In
particular after particle exposure significant delays have been
observed and it has been shown that the extent of delay affects the
expression of damage such as chromosome aberrations. Thus, to
predict how cells respond to ionizing radiation and to derive
reliable estimates of radiation risks, information about
radiation-induced cell cycle perturbations is required. In the
present study we describe and apply
 a method for retrieval of information about the time-course of all cell cycle phases from
 experimental data on the mitotic index only. We study the progression of mammalian cells through
  the cell cycle after exposure. The analysis reveals a prolonged block of damaged cells in the G2 phase.
   Furthermore, by performing an error
analysis on simulated data valuable information for the design of
experimental studies has been obtained. The analysis showed that the
number of cells analyzed in an experimental sample should be at
least 100 to obtain a relative error less than 20\%.
\end{abstract}

\section{Introduction}

It is well-known that cells respond to DNA damage by activating
checkpoints that delay the cell cycle transition in particular from
G1 to S phase and from G2 to M phase. These delays are assumed to
allow additional time for repair (\cite{Li}).

After sparsely ionizing radiation such as X-rays or $\gamma$-rays
mild perturbations of the cell kinetics have been observed
corresponding to the delay of about 1 hour per 1 Gy of exposure
(\cite{Purrott}). In contrast, after particle irradiation dramatic
cell cycle delays have been measured lasting up to 3 cell generation
times (\cite{Scholz_Kraft} and references therein). Moreover, a
rapid desynchronization of initially synchronous cell populations
has been observed after particle exposure (\cite{Scholz_Kraft,Ritter,Ritter2000}). The differences between the effects induced by
both radiation qualities may be directly attributed to spatial
differences in the energy deposition (\cite{Ritter2000,Gudowska}).

Cell cycle delays as observed after radiation exposure are important
for the interpretation of other biological experiments. For example,
recent reports have shown that cell cycle delays interfere with the
time-course of aberrations visible in cells at the first mitosis
post-irradiation. In particular, after particle exposure which
delays the cell cycle progression more than sparsely ionizing
radiation, a drastic increase in the aberration yield with sampling
time has been observed (\cite{Ritter,Ritter2000}) and it has been
recently shown that the average time to enter the first mitosis
correlates directly with the aberration burden of a cell
(\cite{Gudowska}). In other words, cells entering mitosis at later
times harbor more aberrations than those entering mitosis earlier.
Since the frequency of aberrations expressed in first cycle
metaphases is used to determine the absorbed dose (e.g.
\cite{Cucinotta_Durante}) and to derive cancer risk estimates (e.g.
\cite{Mateuca}), the damage sustained by the whole cell population
has to be determined. This can be archived by the analysis of
samples collected at multiple sampling times covering the whole
interval from the first to the last cells reaching mitosis.  Then,
the total yield of aberrations can be determined by a mathematical
approach (i.e. integration analysis, see
\cite{Kaufman,Scholz_Ritter}).

Experimental and theoretical studies of the cell cycle progression
are also of primary interest in the context of cancer therapy
(\cite{Hahnfeldt_Hlatky,Montalenti,Erba,Basse_Ubezio,Wilson}), since
the therapeutic response of solid tumors is known to depend not only
on the administrated dose of ionizing radiation or chemotherapy
agents but also on repopulation and redistribution of cancerous
cells. Cancer therapies may target specific phases of the cell cycle
by blocking or delaying the
progress through one or more phases (see \cite{Montalenti}).\\

The specific aim of our project was to further elucidate the
complexity of particle-induced cell cycle arrest. Investigations
into the effects of particles are becoming increasingly pertinent in
light of rapidly growing interest in this type of radiotherapy
(\cite{Amaldi}). Furthermore, for the planning of manned missions to
Moon and Mars, a better knowledge of the action of charged particles
is needed, since the main contribution of dose during a space
mission outside the magnetic shielding of the Earth originates from
galactic cosmic rays, which are heavy particles from the most
frequent protons to up to iron ions (\cite{Cucinotta_Durante}).

In this study, we present a method for retrieval of information
about all previous phases of the cell cycle, when only the mitotic
index of the experimental cell sample is known. We use the method to
analyze the data measured for mammalian V79 cells after exposure to
Ar ions (\cite{Ritter2000}). V79 cells represent a frequently used
model system to study genotoxic effects of ionizing radiation
(\cite{Ritter,Ritter2000,Weyrather,Groesser,Pathak}) or chemical
agents (\cite{Virgilio}). Information on the cell cycle progression
was available in the form of subsequent measurements of the mitotic
index in control and irradiated samples (Fig. 2) and duration times
of cell cycle phases of control cells (\cite{Sinclair,Scholz}).
Since in the experiment not only the mitotic index, but also the
aberration yield has been measured, the approach may allow us to
study in future the progression of cells carrying a different number
of aberrations.

Mitotic index is the measure of the number of cells undergoing
mitosis at a given time. As the duration of cell cycle phases varies
from cell to cell, the mitotic index depends on the duration
distribution of mitosis, but also on the corresponding duration
distributions of the previous phases. We retrieve these
distributions using a multi-dimensional fit based on general
assumptions concerning the distribution shape, which are inferred
from other experiments \citep{Montalenti}. In this way, having only
the mitotic indices for control and irradiated cells, we were able
to compare their progression through the whole cell cycle and
predict which phases are most vulnerable to irradiation. The fitted
parameters of the duration distributions of cell cycle phases have
also been used to estimate the experimental error of the mitotic
index measurement, depending on the number of cells in the sample.
The estimation has been carried out using the Monte Carlo simulation
of the cell cycle progression for a given number of cells.



\section{Materials and methods}

\subsection{Experimental data used for the analysis }\label{sec:data}

To examine in more detail charged particle-induced cell cycle
progression delays in the first post-irradiation cycle, previously
published data were reanalysed (\cite{Ritter2000}).

For the experiment, V79 Chinese hamster cells were
synchronized by mitotic shake off, a method based on the selective
detachment of mitotic cells from monolayers by shaking. Mitotic
cells were plated in Petri dishes and about 2h after seeding, when
the cells had attached and divided and progressed into G1-phase,
the exposure was done.

 Irradiation with 10.4 MeV u$^{-1}$ Ar ions
(LET = 1226 keV$\mu$m$^{-1}$) was performed with a fluence of
$10^6$ particles cm$^{-2}$ corresponding to a dose of 1.96 Gy,
respectively. At the time of irradiation, at least 95\% of cells
were in G1 phase as determined by flow cytometry. Immediately
after exposure, 5'-Bromo-2'-deoxyuridine (10 $\mu$mg ml$^{-1}$)
was added to the samples to distinguish between metaphases in the
different post-irradiation cycles. Cells were harvested at
multiple sampling times covering the time interval of 2 to 3 cell
generation times. Each sampling time was preceded by a 2h colcemid
treatment (0.1 $\mu$mg ml$^{-1}$) to accumulate mitoses. Since the
main purpose of the experiments was the analysis of the effect of
cell cycle delays on the expression of cytogentic damage,
chromosome preparations were made according to the standard
procedure and slides were stained with the
fluorescence-plus-Giemsa technique. At each sampling time
chromosomal damages was scored in 100 first cycle metaphases
(\cite{Ritter2000}). To gain information on the cell cycle
progression, the mitotic index was determined for each sample by
the direct scoring of 1000-2000 cells on the slides and the cell
generation of at least 200 metaphases was recorded. Due to
experimetal limitation (access to the particle beam and number of
cells that can be synchronized), more detailed measurements of the
cell cycle progression, with a better time resolution,
for example by flow cytometry, were not feasible. A detailed description of the experimental setup is given in (\cite{Ritter2000}).\\

The passage through the cell cycle has been simulated by means of
a kinetic model in which the duration of each phase is taken as a
random variable characterized by its mean and dispersion (see
below). The motivation for such an approach is the rapid
radiation-induced desynchronization of initially synchronous cell
populations. The cell-cycle time (the time interval between cell
divisions) becomes then a random variable whose statistical
properties can be inferred from the analysis of the frequencies of
cells observed in different phases at different times. However, as
mentioned above, the available information on the cell cycle
kinetics relies solely on the measured mitotic indices. Therefore,
for the analysis, these values were used to deduce (a posteriori)
duration times in all phases before the mitosis. In the following
sections, we aim to expand this approach and to present a model
which is general enough to be applied not only for the analysis of
radiation-induced cell cycle delays but also for the analysis of
multi-compartment cell populations perturbed by other cancer
therapies (\cite{Kohandel}).


\begin{figure}[t]

\begin{center}
\epsfig{figure=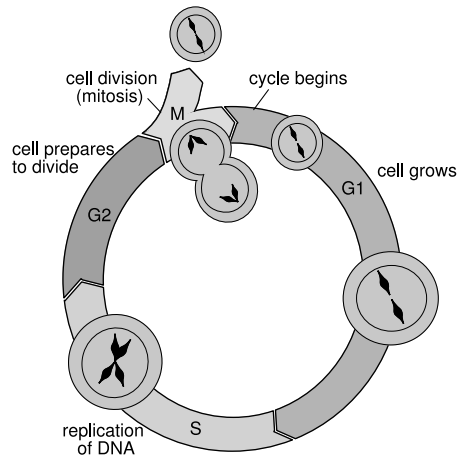, width=6cm}
\end{center}
\caption[]{ \label{fig:cycle} {\small Schematic representation of the cell cycle
 consisting of 4 distinct phases, namely G1, S, G2 and M. For V79
 cells used in the present study the mean duration times of the these
 phases are 2.25 h, 6.5 h, 1.5 h and 0.75h, respectively
 (\cite{Sinclair,Scholz_Kraft}).}}
 \end{figure}


\subsection{Mathematical model of cell cycle progression} \label{sec:model}

In many experiments a positive correlation between radiation dose
and the duration of cell-cycle delays was found. Although such
findings were usually quantified in terms of a linear relationship
between phase duration and dose
(\cite{Zaider_Minerbo,Hahnfeldt_Hlatky}), a more detailed analysis
points to a direct correlation between cell cycle delay and the
number of aberrations carried by a cell (\cite{Gudowska}). This
effect is responsible for the loss of synchrony of the population
and can be illustrated by interpreting the (normalized) mitotic
index as a frequency histogram of times spent by cells before the
actual division happens. The cell cycle kinetics can be then
investigated by treating the durations of the four phases as
independent stochastic variables having probability density
functions described in terms of two adjustable parameters
(\cite{Zaider_Minerbo,Montalenti}). The parameters of such
distributions are fixed in time (stationarity of the distribution is
assumed), whereas the choice of a particular probability
distribution function has been noted not to be
critical for the final result (\cite{Hartmann}).\\

For V79 Chinese hamster cells, as used in our simulations, the
mean duration times of the cell cycle phases are $t_{G1}$ = 2.25
h, $t_S$ = 6.5 h, $t_{G2}$ = 1.5 h, $t_M$ = 0.75h (\cite{Sinclair,Scholz_Kraft}). The phase duration for an individual cell is given
by a certain probability distribution $D_{ph}(\tau)$, where $\tau$
is the time which a given cell had already spent in the current
phase. Consequently, a single cell of phase age $ph$ is assumed to
leave its current phase within a time interval $[\tau, \tau + d\tau]$ with a probability $D_{ph}(\tau)d\tau$. The mean number of
cells $dN_{ph\rightarrow}(t)$ leaving the phase $ph$ within an
infinitesimal time interval $[t, t+dt]$ is defined by the mean
flux $\frac{dN_{ph\rightarrow}(t)}{dt}$.

%
\begin{figure}[t]

\begin{center}

\epsfig{figure=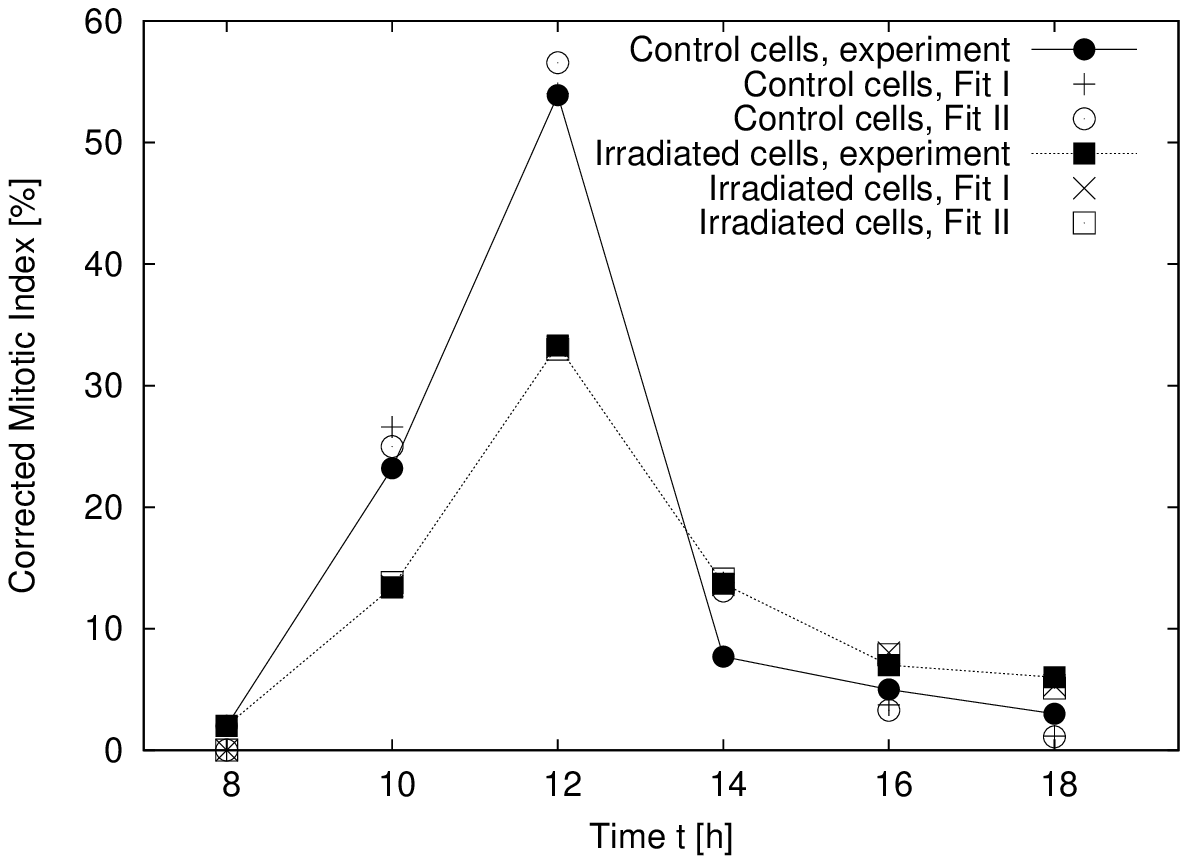, width=9cm}

\end{center}

\caption[]{ \label{fig:mi}{\small Corrected mitotic index of V79 cells exposed in
$G_1$ phase to 1.96 Gy  Ar ions (10.4 MeV/u): experimental data
(\cite{Gudowska, Ritter2000}) along with fits and simulations of
$N_0=1000$ cells are shown. The mitotic index curve was fitted
with a lognormal probability density function with parameters
$\mu$ and $\sigma$ (cf. Eqs.(\ref{eq:fph}, \ref{eq:e})) and the quality of the fit was
determined by using the standard $\chi^2$ test which yielded
almost the same results in two cases. However, the underlying cell
cycle kinetics is quite different for Fit 1 and Fit 2 (see Sec.
\ref{sec:results} and Figs. \ref{fig:lognorm},
\ref{fig:lognorm_rad}). In Fit 1, the G1 phase is largely
dispersed whereas other phases have well defined duration times of
negligible variance. In Fit 2 the phases S and G2 are dispersed
and G1 and M duration times are assigned constant values.}}
\end{figure}


The experimental input to our model is given by the mitotic
indices scored for exposed and control cell populations at
subsequent 2h intervals (\cite{Ritter2000}). Experimentally
measured values of the mitotic index $MI(i)$ at a timestep $i$
\be\label{MI} MI(i) = \frac{N(i) - N(i-1)}{N(i-1)} \ee are defined
as the increase in the total number of cells in relation to their
number at the previous sampling time. Since the time until all
cells reached the first mitosis was quite long compared to the
average cycle length (Fig. \ref{fig:mi}), changes in the
population size due to cell division have to be taken into account
(\cite{Kaufman,Scholz_Ritter}).\\

\begin{figure}[t]

\begin{center}

\epsfig{figure=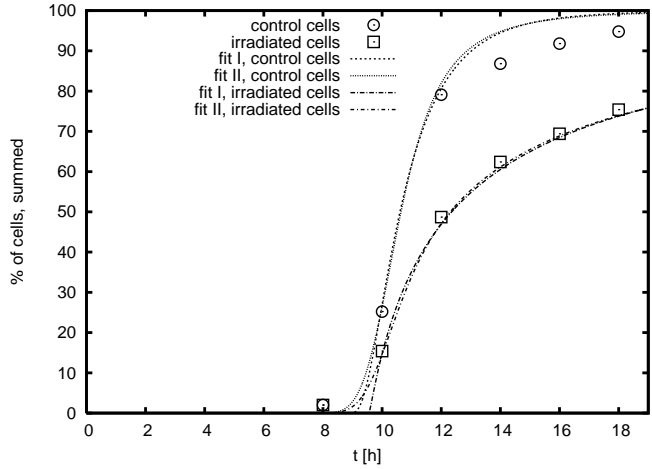, width=9cm}

\end{center}
\caption[]{ \label{fig:int}{\small Cumulative fraction of cells which left the $M$
phase of the first cycle, compared with the experimental results.
For the explanation of fits, see Sec. \ref{sec:results}}}

\end{figure}

To determine the increase in the number of cells in relation to
their initial number one expresses then the corrected mitotic index
in the form \be \label{eq:MI_corr} MI_{corrected}(i) = MI(i)
\frac{N(i-1)}{N(0)} = \frac{N(i) - N(i-1)}{N(0)} \ee In turn, the
summed mitotic indices up to the time step $n$ \be
 S(n)=\sum_{i=0}^n MI_{corrected}(i) = \frac{N(n)-N_0}{N_0}
\ee
give then the fraction of cells which have already completed a full
cycle, i.e. went through mitosis up to the given time. The
experimental results show that this quantity tends to a constant
level, which yields approximately $100\%$ for control cells and less
than $100\%$ for irradiated cells, as displayed in Fig.
\ref{fig:int}.

%


The mean flux $\frac{dN_{ph\rightarrow}(t)}{dt}$ defines a mean
number of cells $dN_{ph\rightarrow}(t)$ leaving the phase $ph$
within an infinitesimal time interval $[t, t+dt]$. The term "mean"
is understood here as an average over a large number of identical
experiments, each with an initial number of $N_0$ cells. Note that
$t$ stands here for the "absolute" time, different from the phase
age $\tau$. In order to relate the description in terms of a mean
flux $\frac{dN_{ph\rightarrow}(t)}{dt} $ based on the ensemble
approach to the description of the time evolution of a single cell
whose probability of leaving the current phase is $D_{ph}(\tau)$,
we assume that we deal with a large ensemble of samples, each
initially containing $N_0$ cells in the same phase $ph$. Moreover,
the sample populations are assumed to be of the same phase age,
e.g. $\tau=0$. We start our "numerical experiment" at time $t=0$,
so that $t=\tau$, and count the cells which leave the phase within
the age $\tau$ and $\tau+d\tau$. Their mean number will be given
by:

\be
dN_{ph\rightarrow}^{(1)}(\tau)=N_0 F_{ph}(\tau)d\tau,
\ee

where $F_{ph}(\tau)$ is a probability distribution which gives the
likelihood of a cell leaving the phase at
the time $t = \tau$.\\

On the other hand, if we trace the behavior of a particular cell,
we can define $D_{ph}(\tau)$ as the conditional probability distribution of
leaving the phase at age $\tau$, provided that the cell had not
completed the phase earlier:

\be\label{eq:D} D_{ph}(\tau) =\frac{ F_{ph}(\tau)}{1-\int_0^\tau
F_{ph}(\tau')d\tau'}. \ee

Function $F_{ph}(t)$ in the above formula represents the
distribution of phase duration times in an initially synchronized
population of cells. We follow the suggestions of Zaider and Minerbo
(1993) and Montalenti et al. (1998) and assume a lognormal
distribution of phase duration (\cite{Engen_Lande}):

\be \label{eq:fph}
 F_{ph}(t)=\frac{1}{t\sqrt{2 \pi \sigma_{ph}^2}}\exp\left(-\frac{(\ln{t}-\mu_{ph})^2}{2\sigma_{ph}^2}\right).
\ee

with parameters $\mu_{ph}$ and $\sigma_{ph}$ determining the mean

\be \label{eq:e} E[t]=e^{\mu_{ph}+\frac{\sigma^2_{ph}}{2}} \ee

and variance

\be
E[(t-E[t])^2]=e^{2\mu_{ph}+\sigma^2_{ph}}(e^{\sigma_{ph}^2}-1).
\ee

In the above expressions, $\mu$ and $\sigma$ are the parameters to
be fitted.

It should be noticed that a log-normal distribution is not the only
possible choice to be postulated as suitable for the description of
the random distribution of phase duration times. The other
possibility might be, e.g., the gamma distribution resulting from a
mixture of exponential distributions expected in simple renewal
processes (\cite{Feller,Zaider_Minerbo,Hahnfeldt_Hlatky}).

Ensemble analysis of the mean flux allows splitting
$dN_{ph\rightarrow}(t)$ into a sum of contributions from subsequent
cycles: \be \label{dN}
dN_{ph\rightarrow}(t)=\sum_{i=1}dN_{ph\rightarrow}^{(i)}(t) \ee
$dN_{G1\rightarrow}^{(1)}(t)$ is defined as the product of the
probability of leaving the $G1$ phase within the time interval
$(t,t+dt)$ and the mean total number of cells at time $t$: \be
dN_{G1\rightarrow}^{(1)}(t)=dP_{G1\rightarrow}(t)N(t), \ee which
means that \be dP_{G1\rightarrow}(t) = F_{G1}(t)dt. \ee In the first
cycle $N(t)=N_0$, so \be
dN_{G1\rightarrow}^{(1)}(t)=dP_{G1\rightarrow}^{(1)}(t)N_0= N_0 dt
F_{G1}(t). \ee The probability of leaving the $S$ phase in the first
cycle at a certain time will depend on the probability of leaving
the previous phase, and thus \be
dN_{S\rightarrow}^{(1)}(t)=dP_{S\rightarrow}^{(1)}(t)N_0= N_0
dt\int_0^tF_{G1}(t-\tau)F_S(\tau)d\tau= N_0 dt F_1(t), \ee and,
analogously: \be
dN_{G2\rightarrow}^{(1)}(t)=dP_{G2\rightarrow}^{(1)}(t)N_0= N_0 dt
\int_0^t F_1(t-\tau)F_{G2}(\tau)d\tau= N_0 dt F_2(t), \ee \be
\label{eq:Mout}
dN_{M\rightarrow}^{(1)}(t)=dP_{M\rightarrow}^{(1)}(t)N_0= N_0 dt
\int_0^t F_2(t-\tau)F_{M}(\tau)d\tau= N_0 dt F_3(t). \ee In the
moment of leaving the M phase, cells divide: two G1 cells are
produced and therefore, the number of cells in G1 phase (and thus
the number of those leaving G1 phase) is twice larger: \be
dN_{G1\rightarrow}^{(2)}(t)=2dP_{G1\rightarrow}^{(2)}(t)N_0=2 N_0 dt
\int_0^t F_3(t-\tau)F_{G1}(\tau)d\tau= 2 N_0 dt F_4(t). \ee In this
manner we generate consecutive contributions to Eq. (\ref{dN}).
Knowing the mean flux as given by Eq. (\ref{dN}) we can calculate
other mean quantities of interest, such as mean number of cells
found in the phase $ph$ at a given time after irradiation. This
method gives therefore a global view on the cell cycle kinetics.

In order to reconstruct the fate of an individual cell, we have used
the Monte Carlo method. The computer simulation tracks the cycle of
each single cell separately. Initially we generate an ensemble of
$N_0$ cells. Each cell is assumed to start in the $G1$ phase and its
phase age is $\tau=0$. In subsequent time steps each cell goes
through its individual cycle, and in every time step $d\tau$ each
cell has a probability $D_{ph}(\tau)d\tau$ to leave its current
phase.
T
he algorithm is built based on the following scheme:\\
\noindent 1. The cell is in the phase $ph$. Its phase age is $\tau$.\\
2. A random number $x$, $0 \leq x<1$ is generated.\\
3. If $x<D_{ph}(\tau)d\tau$, the programme proceeds to 4., else it goes to 6.\\
4. If the next phase $"ph+1"\neq G1$, then the cell exits from its
current phase:
 $ph="ph+1"$, $\tau=0$. Go to 1.\\
5. If the next phase $"ph+1"=G1$, then the cell exits from its
current phase and replicates: $ph=G1$, $\tau=0$. Additionally, a new
cell is generated, whose current phase is $G1$ and
$\tau=0$. The cycle starts with going to 1.\\
6. The cell remains in its current phase: $\tau=\tau+d\tau$.
At the next step the programme starts with 1.\\

All integrations to compute $D_{ph}(\tau)$ from (\ref{eq:D}) have
been performed in our simulation using the Euler method with discrete time steps $\Delta t$.\\

The above described approach allows to examine the kinetics of a
single experiment. Computing the variance of quantities of interest,
we can estimate how much a single realization of the experiment
deviates from the ensemble average.


%
 \begin{figure}[t]
\begin{center}
\epsfig{figure=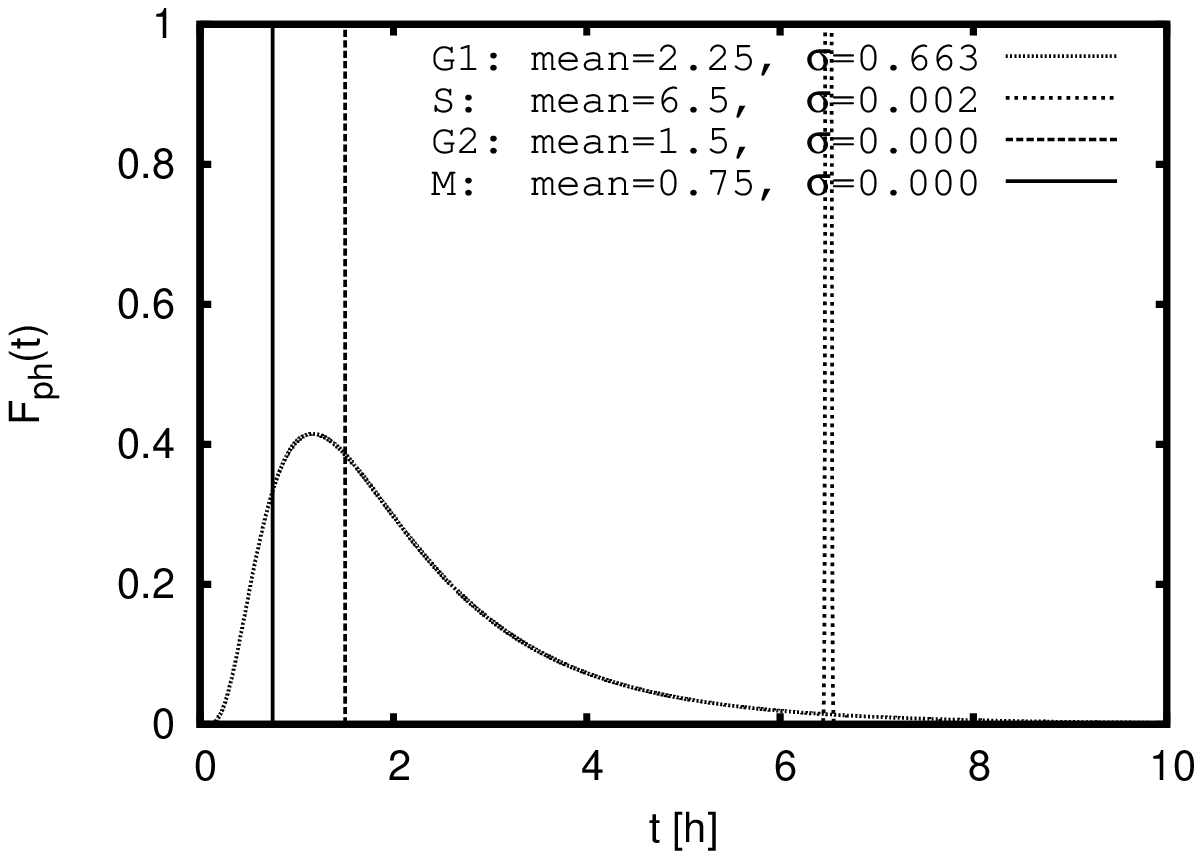, width=9cm} \end{center}
\begin{center}
\epsfig{figure=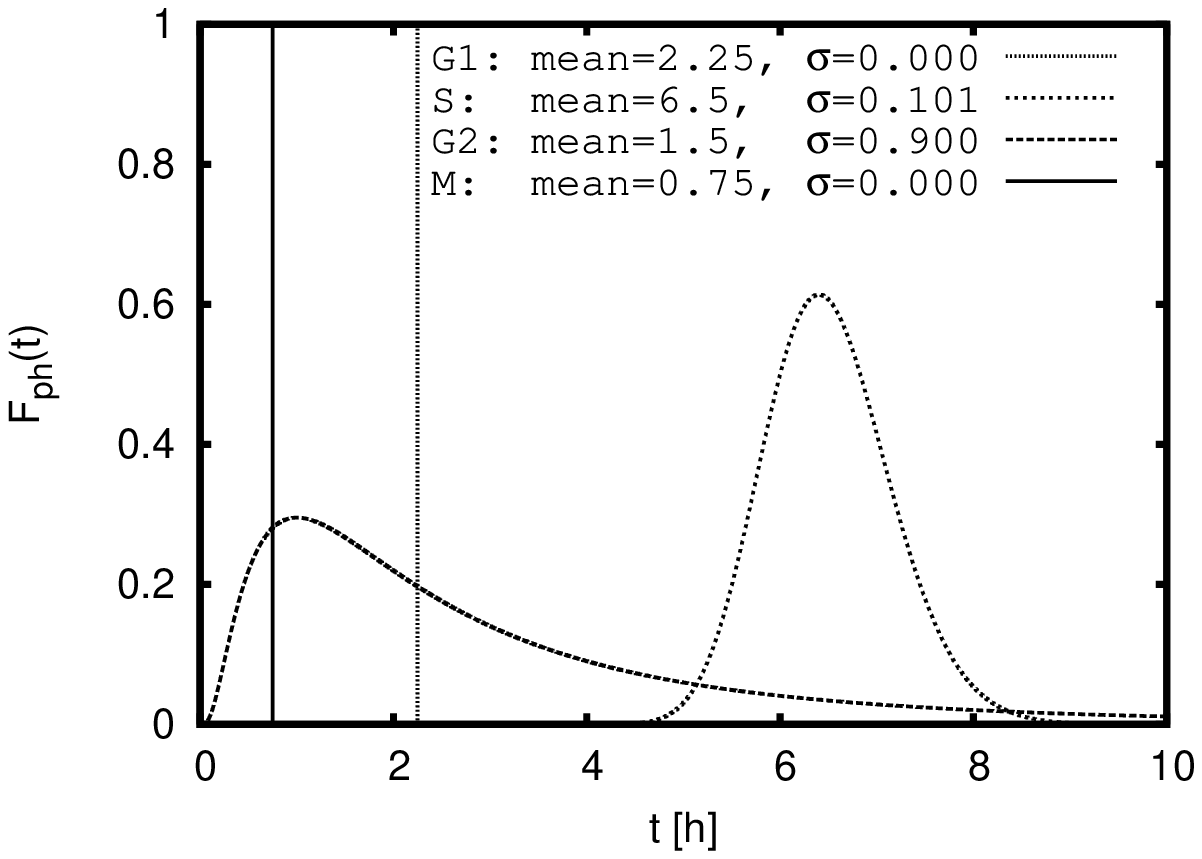, width=9cm}
\end{center}
 \caption[]{ \label{fig:lognorm}{\small Control cells: Probability density functions for
 leaving a phase in phase age $\tau$ (top panel: Fit 1, bottom
 panel: Fit 2) }}
 \end{figure}


%
\begin{figure}[t]
\begin{center}
\epsfig{figure=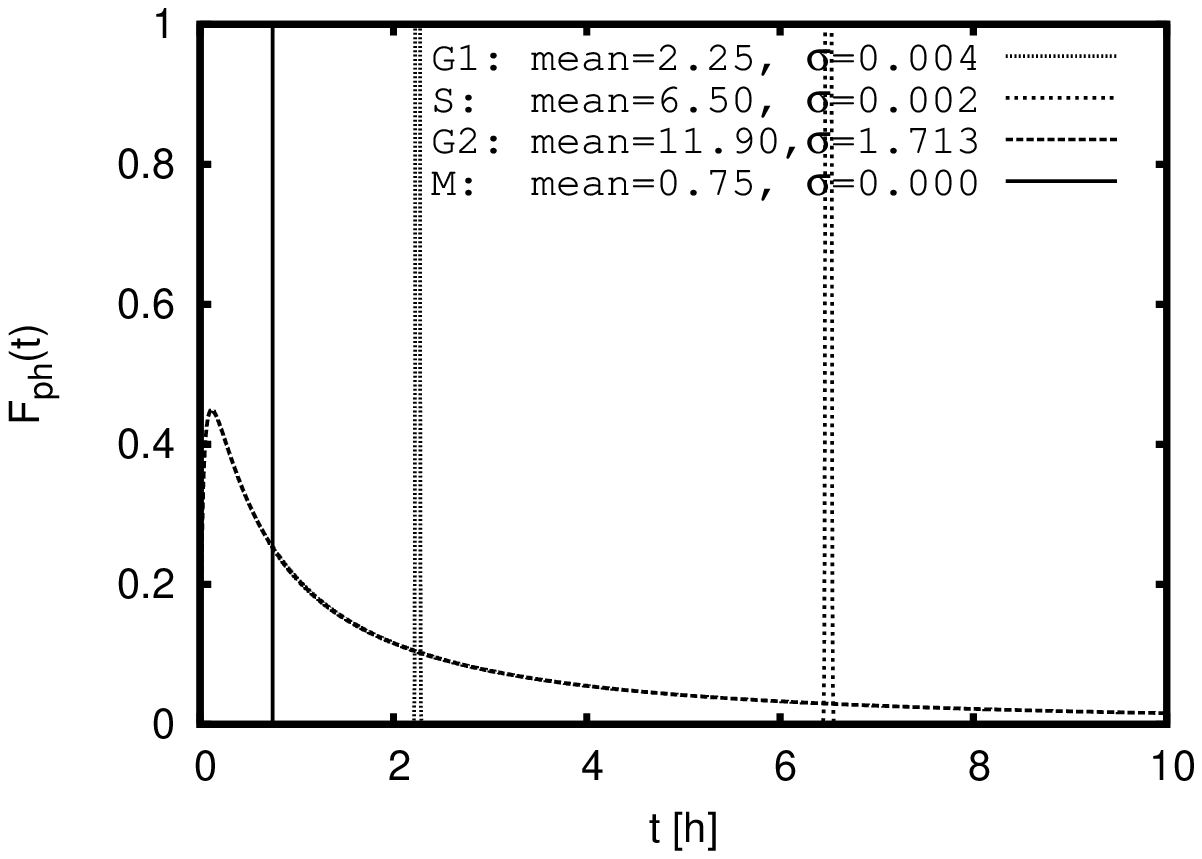, width=9cm} \end{center}
\begin{center}
\epsfig{figure=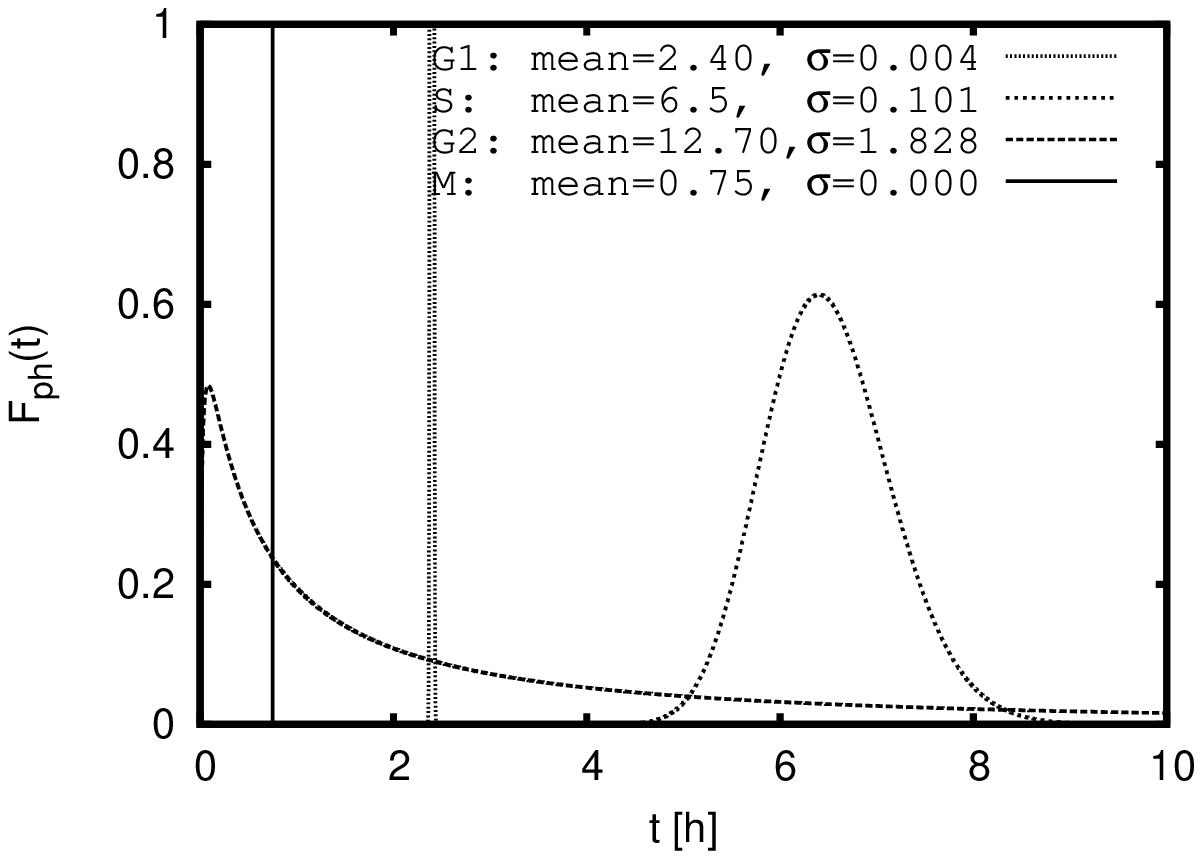, width=9cm}
\end{center}
 \caption[]{ \label{fig:lognorm_rad}{\small Irradiated cells: Probability density functions
 for leaving a phase in phase age $\tau$ (top panel: Fit 1, bottom
 panel: Fit 2), cf. Eqs. (\ref{eq:fit1}, \ref{eq:fit2}).}}
 \end{figure}

\begin{table}[t]
\begin{tabular}{|l|l|l|l|l|l|l|}
\hline
\multicolumn{7}{|c|}{Control cells} \\
\hline
\multicolumn{1}{|c|}{} & \multicolumn{3}{c|}{Fit I} & \multicolumn{3}{c|}{Fit II} \\
\hline
Phase & Median & Mode  & CV    & Median & Mode  & CV    \\
\hline
G1    &  1.806 & 1.164 & 1.346 &  2.25  & 2.25  & -     \\
S     &  6.500 & 6.500 & 500   &  6.467 & 6.401 & 9.876 \\
G2    &  1.5   & 1.5   & -     &  1.000 & 0.445 & 0.895 \\
M     &  0.75  & 0.75  & -     &  0.75  & 0.75  & -     \\
\hline
\multicolumn{7}{|c|}{Irradiated cells} \\
\hline
\multicolumn{1}{|c|}{}& \multicolumn{3}{c|}{Fit I} & \multicolumn{3}{c|}{Fit II}\\
\hline
Phase & Median & Mode  & CV    & Median  & Mode  & CV   \\
\hline
G1    &  2.250 & 2.250 & 250   &  2.398  & 2.398 & 250  \\
S     &  6.500 & 6.500 & 500   &  6.467  & 6.401 & 9.876\\
G2    &  2.744 & 0.146 & 0.237 &  2.389  & 0.084 & 0.192\\
M     &  0.75  & 0.75  & -     &  0.75   & 0.75  & -    \\
\hline
\end{tabular} 
\caption{\label{tab:mode_cv} Median, modal values and coefficients of variation for the fitted phase duration time distributions $F_{ph}$.}
\end{table}

\section{Results} \label{sec:results}
Using the methods described above, we have first fitted the
experimental results, i.e. the mitotic indices (see Sec.
\ref{sec:data}). Since the experimentally accessible information
refers to cells initially synchronized in G1 phase, we assumed an
"ideal synchrony" for the numerical analysis , i.e. at time $t=0$
all of cells were assumed to be of the phase age $\tau=0$.

\subsection{Control cells}\label{subs:control}
The first step in our analysis was fitting the parameters of the
$F_{ph}$  distributions (\ref{eq:fph}) for each phase, in such a way
that they together made up the time-course of mitosis
(\ref{eq:Mout}) which fitted the experimental corrected mitotic
indices (\ref{eq:MI_corr}).

We performed the integrations numerically with a time-step $\Delta t=0.03$ h (Euler method), obtaining an approximate flux:
\be
\frac{\Delta N_{ph\rightarrow}(t)}{\Delta t} \approx
\frac{dN_{ph\rightarrow}(t)}{dt}.
\ee
Having evaluated the outflow
from the M phase, we could further easily compute the total mean
number of cells at a given time, from which the (mean) corrected
mitotic index (\ref{eq:MI_corr}) was constructed. (We recall that
the word "mean" is understood here as an average over a large
number of
identical experiments.)\\

Since
we would have had to fit up to 8 parameters, we decided to
simplify the task. The results cited in Sec. \ref{sec:data}
delivered the values of mean duration times of each phase:
$\overline{t}_{G1}=2.25$ h, $\overline{t}_{S}=6.5$ h,
$\overline{t}_{G2}=1.5$ h, $\overline{t}_{M}=0.75$h.
Therefore, we used them as the four (frozen) parameters needed for
modelling and focused on four others (we chose $\sigma_{ph}$,
which can be treated as a measure of fitted distribution's width)
which had to be chosen by the best fit. The initial number of
cells assumed for the modelling was $N_0=1000$ and the parameters
were fitted by using a Metropolis algorithm.\\

We found that the result of the fitting (see Fig.
\ref{fig:mi}) was ambiguous and there were two possible
least-square fits with $\chi^2$ value of the same order. The
inferred coefficients displayed a significant variability between
both data sets (see Fig. \ref{fig:lognorm} and Tab. \ref{tab:mode_cv}):\\

Fit 1:
\be \label{eq:fit1}
\sigma_{G1}=0.663, \sigma_{S}=0.002, \sigma_{G2}=0.000, \sigma_{M}=0.000, \chi^2=55.0
\ee

Fit 2:
\be \label{eq:fit2}
\sigma_{G1}=0.000, \sigma_{S}=0.101, \sigma_{G2}=0.900, \sigma_{M}=0.000, \chi^2=50.0
\ee

Although the two possible cell cycle schemes produced almost
identical mitotic index curves, the intra-cycle distributions
differed very much. In Fit 1, the G1 phase is largely dispersed
whereas other phases have well-defined (deterministic) duration
times. In contrast, in Fit 2, S and G2 phases are dispersed with
the duration times of G1 and M phases being well specified.


%
\begin{figure}[t]
\begin{center}

\epsfig{figure=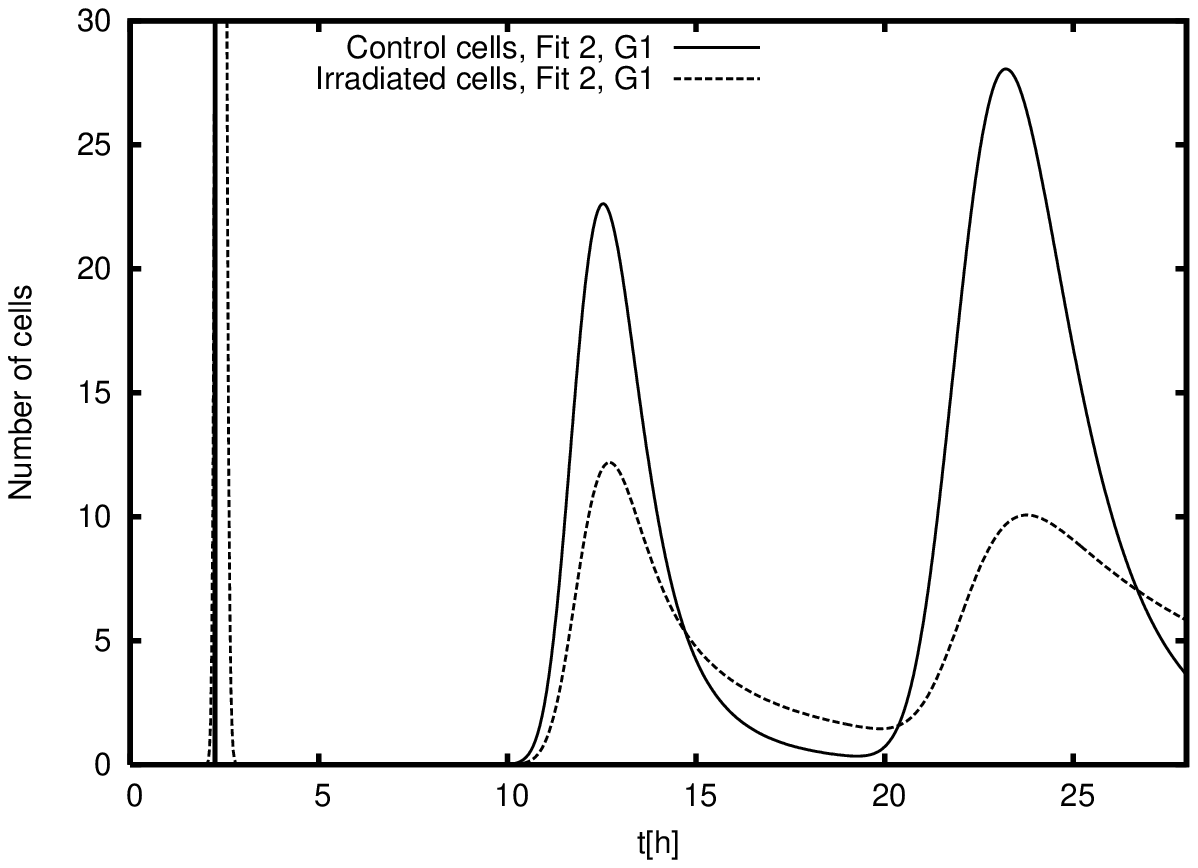, width=9cm}

\end{center}

\begin{center}

\epsfig{figure=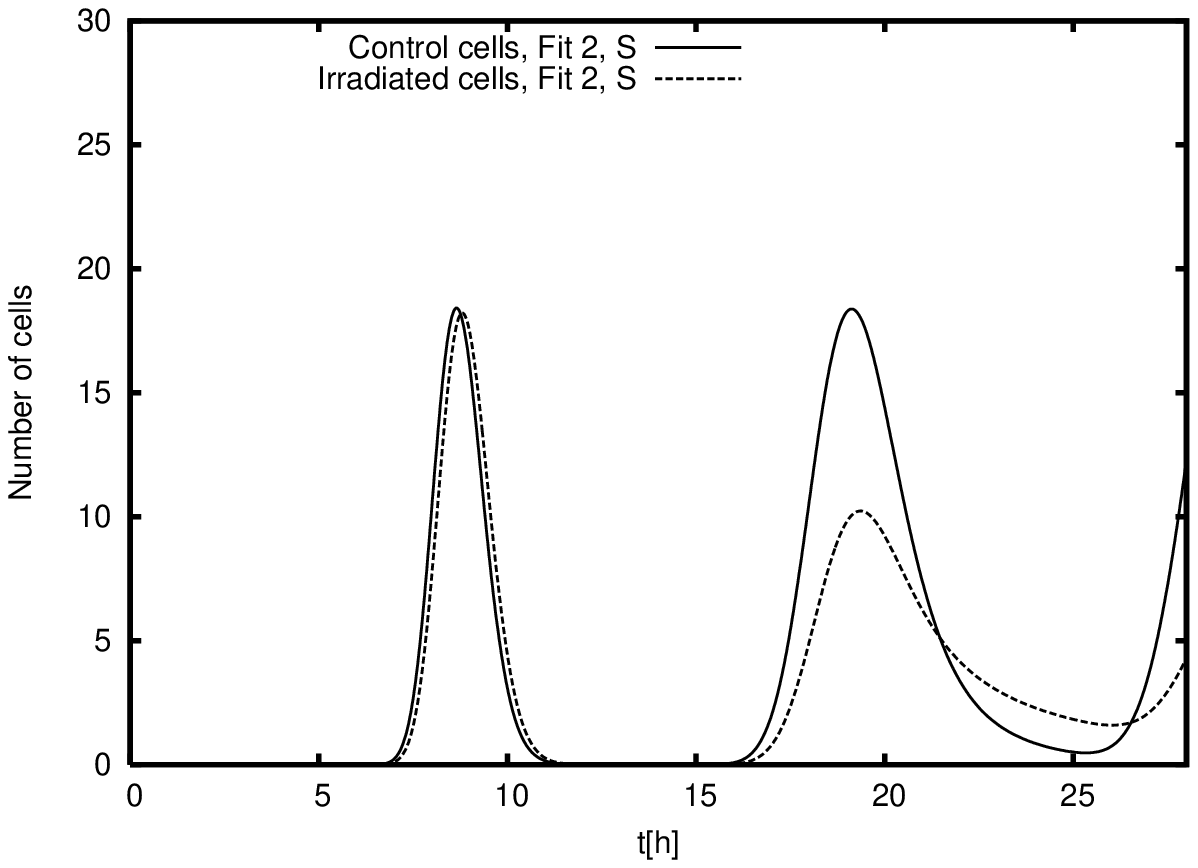, width=9cm}

\end{center}
\caption[]{ \label{fig:flux}{\small Flux from phases G1 and S (number of cells passing per 1h) at time $t$, compared
for control and irradiated cells (Fit 2). }}
\end{figure}

 \begin{figure}[t]
\begin{center}
\epsfig{figure=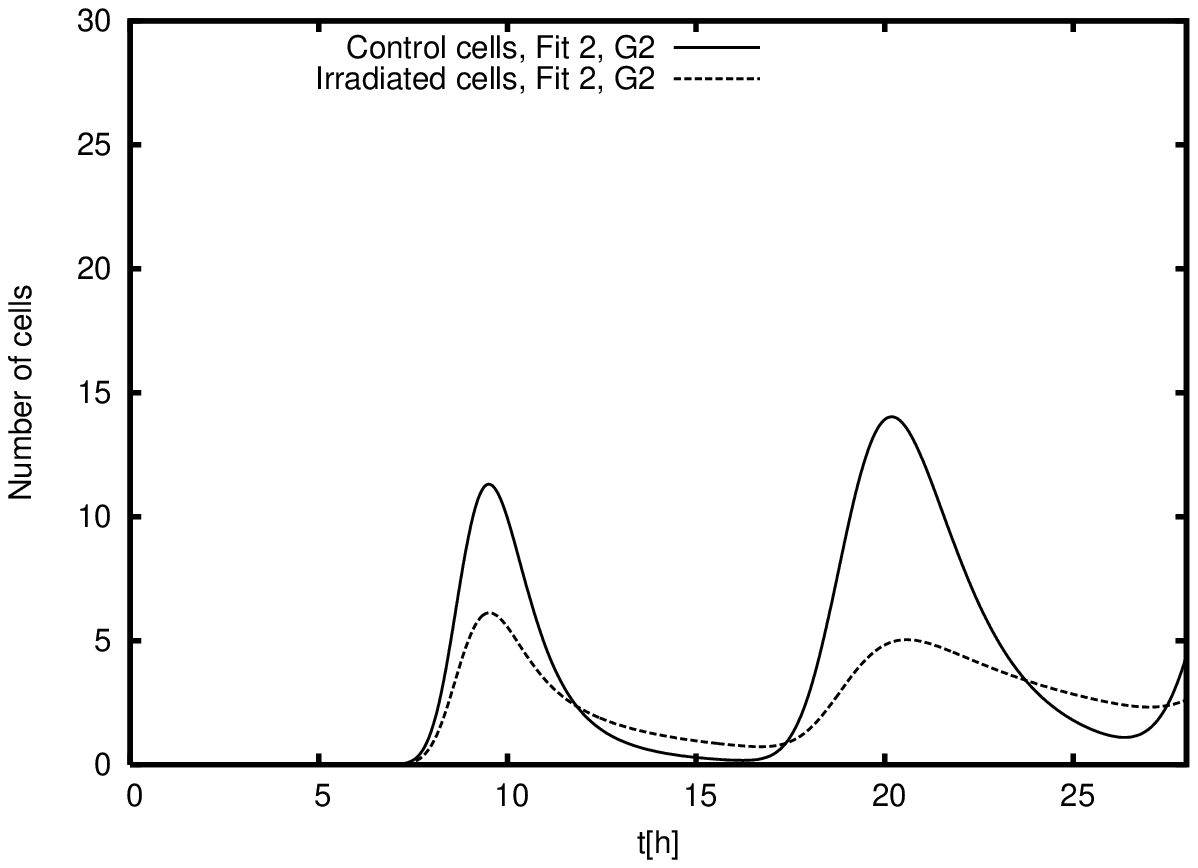, width=9cm}
\end{center}
\begin{center}
\epsfig{figure=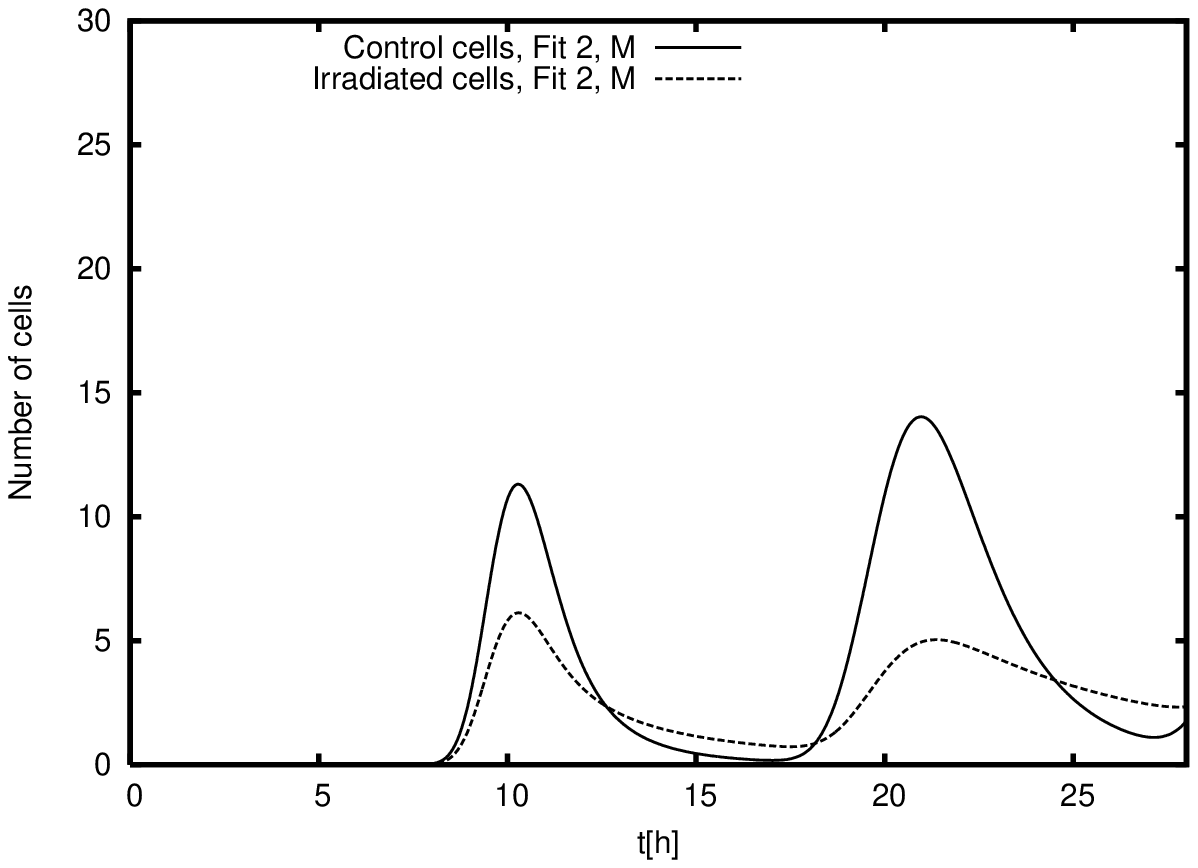, width=9cm}
\end{center}
 \caption[]{ \label{fig:flux1}{\small Flux from phases G2 and M (number of cells passing per 1h) at time $t$, compared
 for control and irradiated cells (Fit 2). } }
 \end{figure}


\subsection{Irradiated cells}
In order to reproduce the mean corrected mitotic indices of the
irradiated cell population we performed a similar fitting procedure
as mentioned above. To proceed, we made the biologically justified
assumption that irradiation affects only the characteristics of G1
and G2 phases, since checkpoints in G1 and G2 are known to block the
cell cycle progression to give the cell time to repair  and continue
cycling or to undergo apoptosis (\cite{Li}). We also expected that
the mean duration times of both phases, as well as the variances of
the duration times should increase because the damaged cells would
be blocked in either one of them for a longer time. Therefore, we
fitted only the parameters $\overline{t}_{G1}, \sigma_{G1},
\overline{t}_{G2}, \sigma_{G2}$, taking others as fixed and
referring to their values obtained in the former fit for control
cells. While doing so, we did not introduce any additional mortality
parameter. Instead, we assumed solely that
the damaged cells will stay in G1 or G2 phase for a very long time.\\

\noindent For the parameters corresponding to Fit 1 the analysis
yielded:

\be \label{eq:fitrad1}
\overline{t}_{G1}=2.25,
\sigma_{G1}=0.004, \overline{t}_{G2}=11.90, \sigma_{G2}=1.713,
\chi^2=5.6,
\ee

whereas  for Fit 2 it resulted in:

\be\label{eq:fitrad2}
\overline{t}_{G1}=2.40, \sigma_{G1}=0.004,
\overline{t}_{G2}=12.70, \sigma_{G2}=1.828, \chi^2=5.8
\ee

In both fits the $F_{G2}(\tau)$ distribution function became wider
and its mean duration time strongly increased (Fig. \ref{fig:lognorm_rad}, see also Tab. \ref{tab:mode_cv}).

In turn, the $F_{G1}(\tau)$ distribution in Fit 1 became narrower
in comparison to the corresponding distribution for the control
cells, which implied that Fit 1 was unrealistic. This finding is consistent with previous flow cytometric studies showing that V79 cells suffer short cell cycle delays in
G1 phase, but more pronounced delays in S- or G2-phase (Scholz et
al. 1994). The low activity of the G1 checkpoint in V79 cells
might result from mutations in the p53 gene (Chaung et al,
1997). We have therefore concluded that  Fit 2 was confirmed as better
matching the biological scenario: here the mean duration time of
G1 increased and the distribution function became wider. The very
long tail of $F_{G2}(\tau)$, as predicted in this case,  suggests
that damaged cells may remain blocked in the G2 phase for an
extremely long time with a non-zero probability
as it has been experimentally shown by Scholz et al.
(1994).

%
 \begin{figure}[t]

\begin{center}

\epsfig{figure=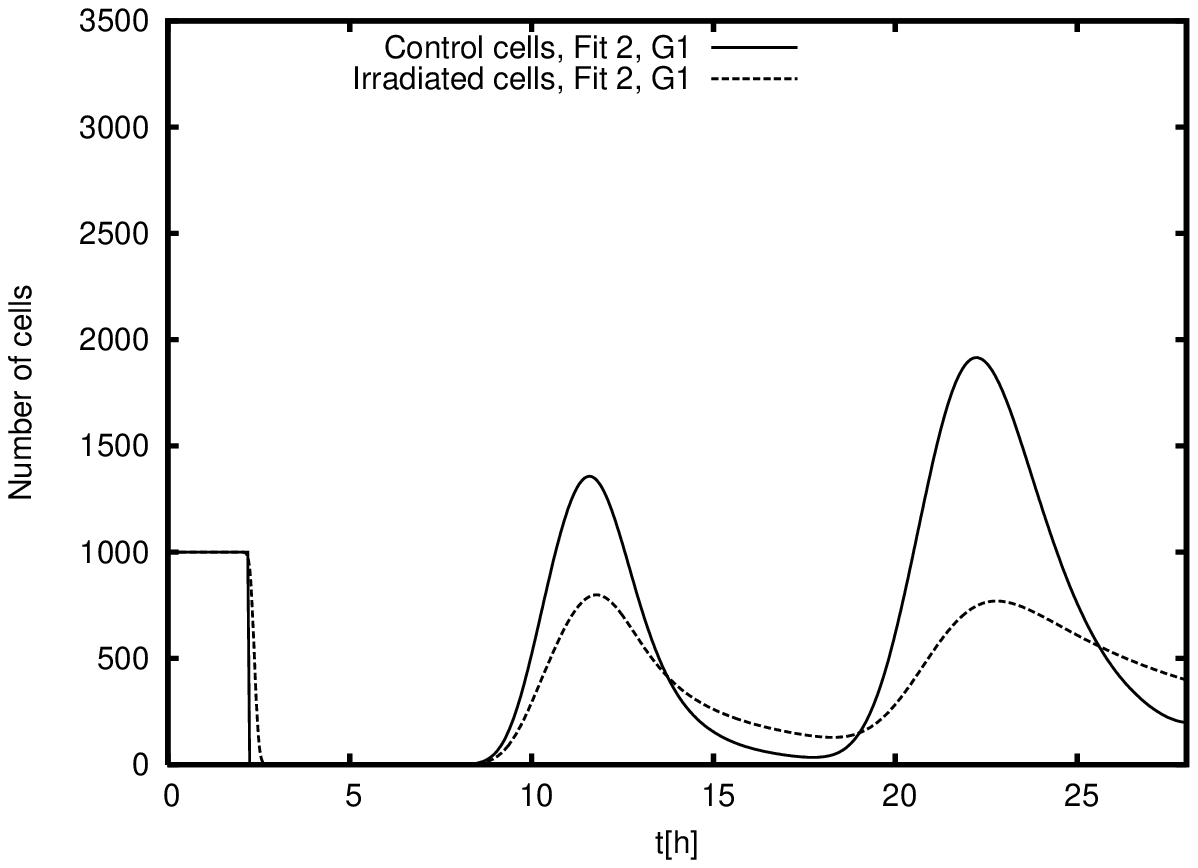, width=9cm} \end{center}

\begin{center}

\epsfig{figure=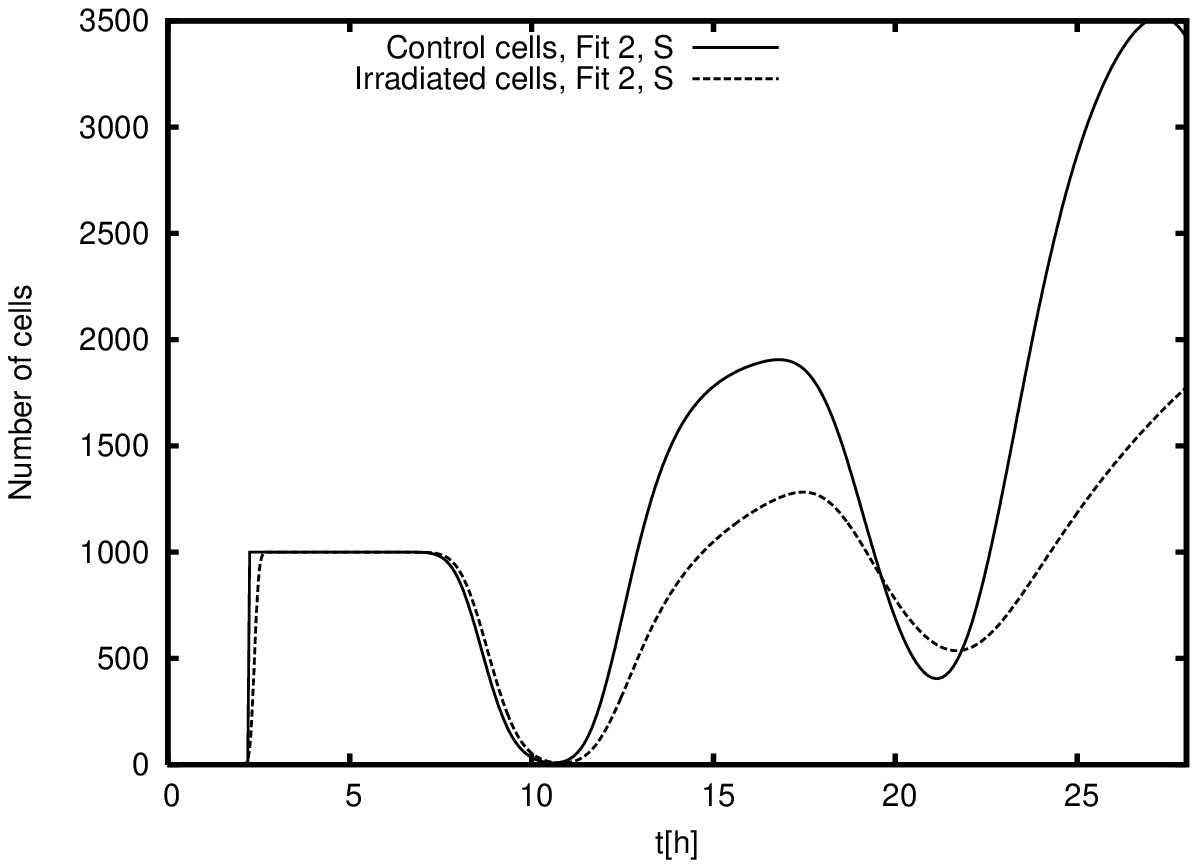, width=9cm}

\end{center}

 \caption[]{ \label{fig:number}{\small Number of cells in phases G1 and S at
 time $t$, compared between control and irradiated cells (Fit 2). }}

 \end{figure}


%
\begin{figure}[t]

\begin{center}

\epsfig{figure=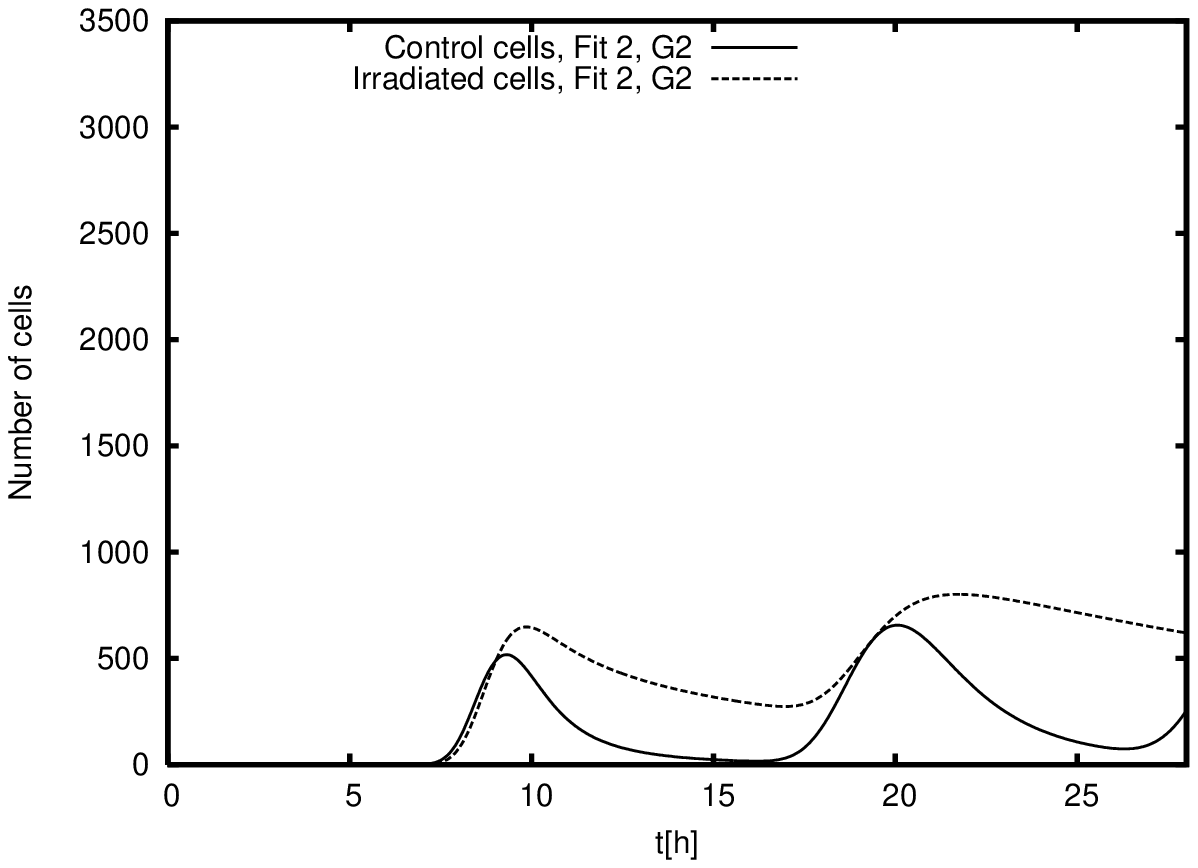, width=9cm} \end{center}

\begin{center}

\epsfig{figure=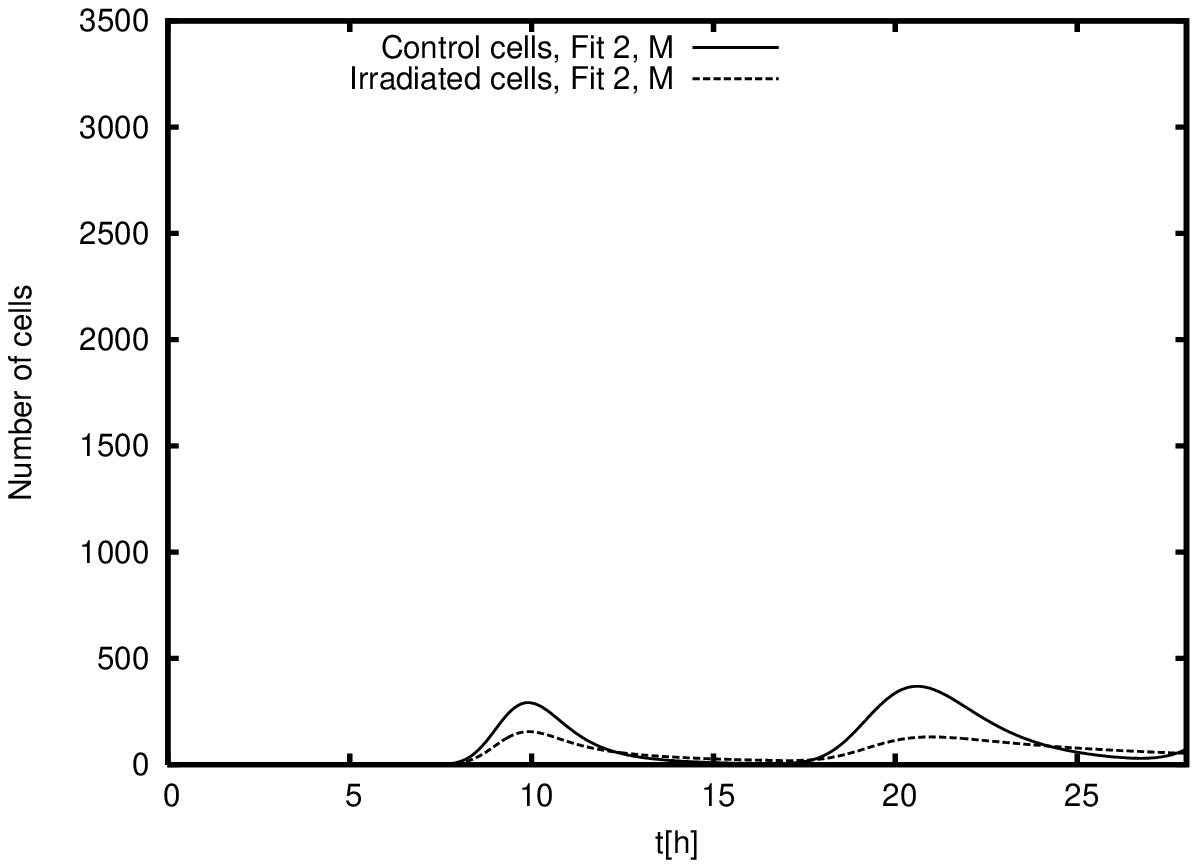, width=9cm}

\end{center}

\caption[]{ \label{num1}{\small Number of cells in phases G2 and M at
time $t$, compared between control and irradiated cells (Fit 2). }}
\end{figure}


Figs.~\ref{fig:flux} and \ref{fig:flux1} compare the progression of
cell fluxes between subsequent phases for control and irradiated
populations, respectively, while the number of cells remaining in
a given phase at time $t$ is shown in Figs. \ref{fig:number} and \ref{num1}. In the first
cell cycle a notably slower outflow of cells from the G2 phase is
visible which corroborates with the number of cells remaining in
that phase up to 16h after exposure. Cells blocked in that phase
also diminish the fraction of those which are able to enter
mitosis up to the same time-point. As expected, the cumulative
number of cells which enter mitosis is much lower in the
irradiated population than in the control sample and can be well
correlated with the fraction of cells arrested in G2.


%
 \begin{figure}[t]
 \begin{center}
 \epsfig{figure=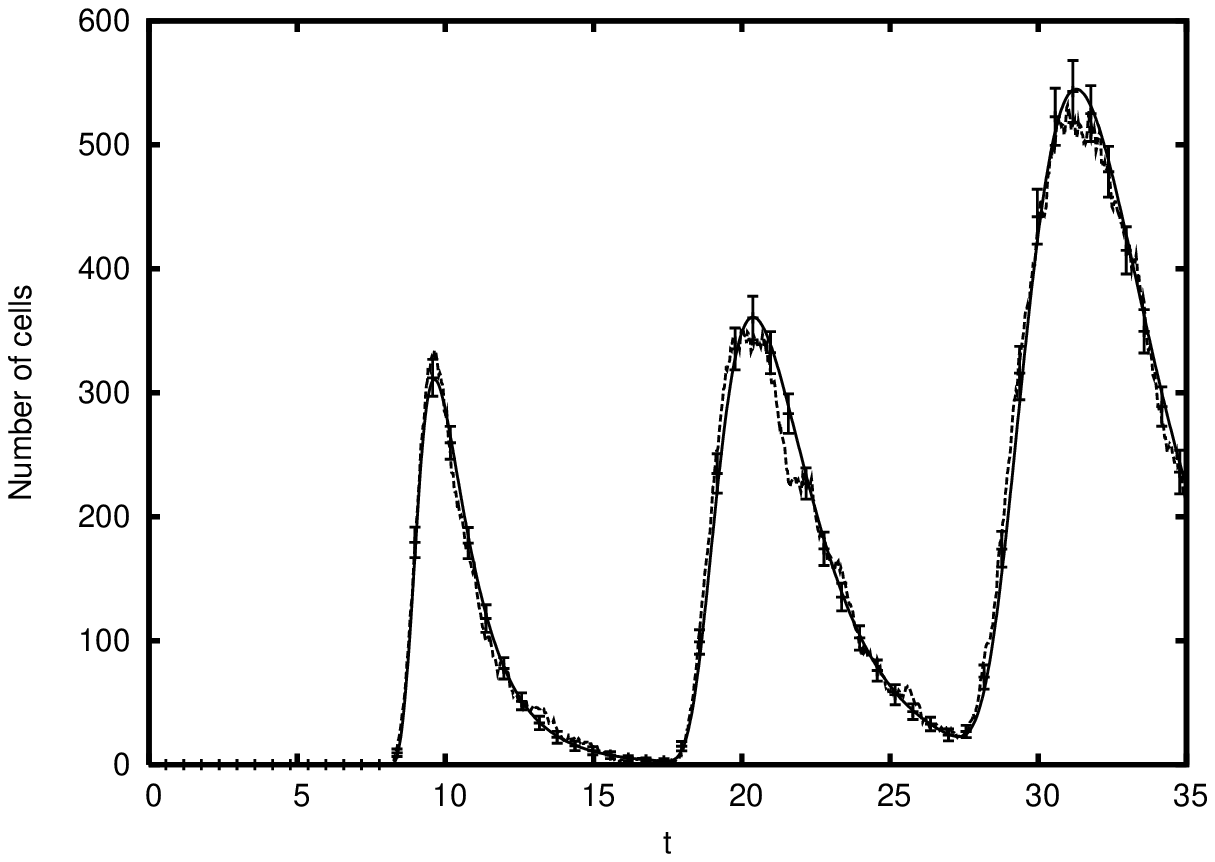, width=9cm}
 \end{center}
 \caption[]{  \label{fig:simM}{\small Number of cells found in M phase at time $t$.
 Black line: theoretical prediction obtained using the mean flux
 approach. Dashed line: a single simulation for $N_0=1000$ cells. Error
 bars: standard deviation for the simulation.}}
 \end{figure}


%
 \begin{figure}[h]
\begin{center}
\epsfig{figure=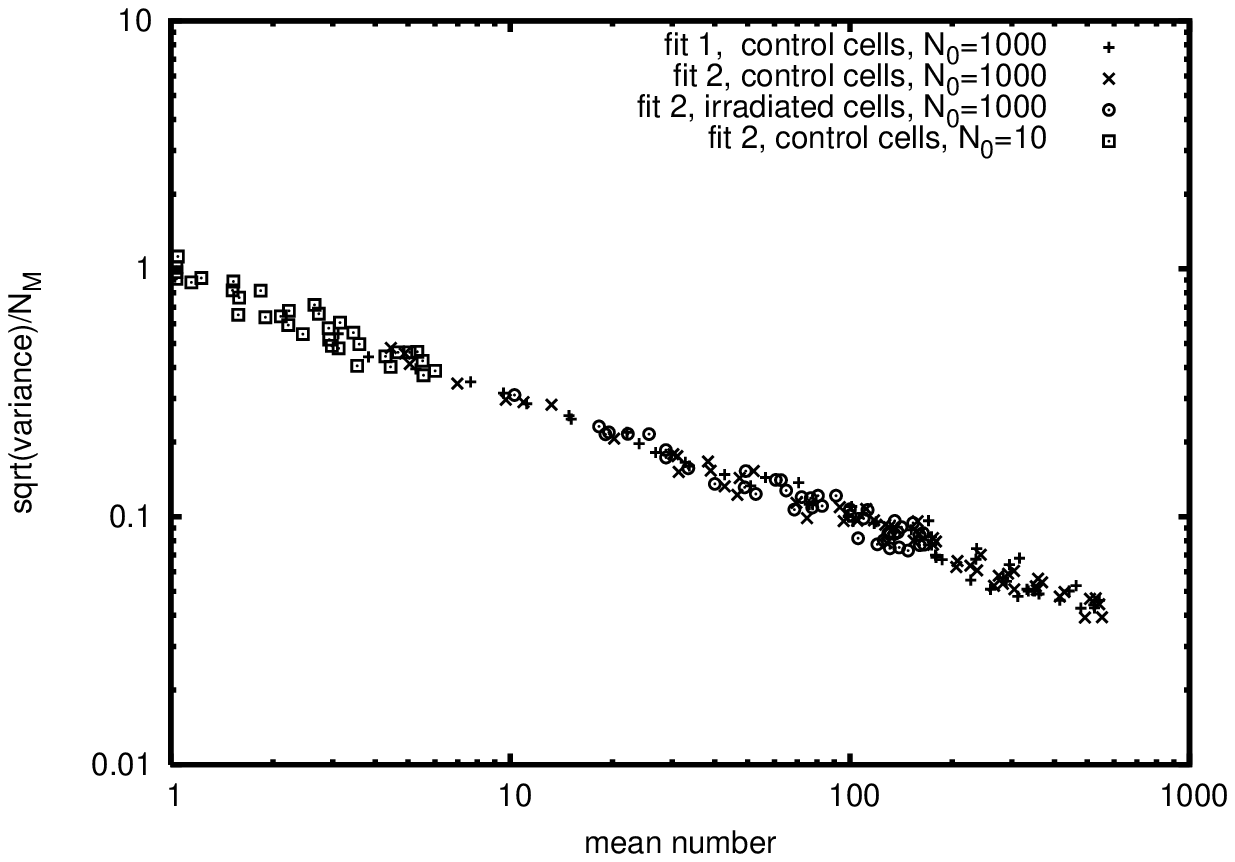, width=9cm}

\end{center}
 \caption[]{ \label{fig:error2}{\small Relative error of the measurement vs. mean total
 number of cells analyzed. }}
 \end{figure}


\subsection{Determination of experimental errors}

Another important finding of our work is that the applied approach
allows for estimation of inter-experimental differences resulting
from the stochastic character of phase duration times. By performing
a Monte Carlo simulation, described in Sec. \ref{sec:model} we
simulated a time-evolution of an ensemble of cells (here:
$N_0=1000$) using the set of previously fitted parameters. The
number of cells at different moments of time was counted, and
therefrom the mitotic index (\ref{MI}) was derived. Using this
method, the variance/standard deviation of the corresponding
probability distribution function was
estimated (see Fig. \ref{fig:error2}).\\

We performed a series of $100$ cell cycle simulations with the
initial number of cells $N_0$. The mean number of cells found in a
given phase $\langle N_{ph}(t)\rangle$ and the variance
$\sigma^2_{N_{ph}}(t)$ were calculated. Analyzing the mutual
dependence of the relative error $\sigma_{N_{ph}} \over \langle
N_{ph} \rangle$ and $\langle N_{ph} \rangle$ we found that the
relative error does not depend on the initial number of cells $N_0$
nor on the set of parameters of $F_{ph}$. It depends only on the
current number of cells. A practical conclusion drawn from these
results is that the number of cells analyzed in an experimental
sample should be at least of order of $100$ to obtain a relative
error less than $20\%$. If a better precision is required, a larger
cell population has to be examined.

\section{Discussion}
In studies preceding this project (\cite{Ritter,
Ritter2000,Scholz_Ritter,Gudowska}) the relationship between
radiation-induced mitotic delay and expression of chromosome damage
was shown. To further establish differences in progression through
the division cycle of unirradiated (control) and exposed cells, we
developed a method for retrieval of the information about the
time-course of all cell cycle phases, when only the mitotic index of
a cell sample is known. The method consists in a multi-dimensional
fit based on general assumptions concerning the duration
distribution of cell cycle phases, which are inferred from other
experiments \citep{Montalenti}.

The analysis of the experimentally measured mitotic indices for
mammalian V79 cells after exposure to Ar ions (\cite{Ritter2000})
revealed a prolonged block of damaged cells in the G2 phase.
Additionally, we have shown that this approach is applicable
directly to determine experimental errors resulting from the
stochastics of phase duration times.

In future studies the quality of the model will be validated through
the application to data sets generated for other cell lines and cell
types such as human skin fibroblasts and lymphocytes whose
cell-cycle kinetics after irradiation is altered differently
(\cite{Flatt,Lee,Tenhumberg}). For example, it has been shown that normal human
fibroblasts exposed in G0/G1 phase to ionizing radiation suffer a
prolonged G1 arrest (\cite{Flatt,Tenhumberg}), while human
lymphocytes are predominantly arrested in G2 (\cite{Lee}). Furthermore, the model may be
used to analyze the progression of cells carrying different numbers of aberrations.\\

\bibliographystyle{spbasic}
\bibliography{cellcycled_rev-SR}

%


\end{document}